\documentclass[baaa]{baaa}

\usepackage[pdftex]{hyperref}
\usepackage{subfigure}
\usepackage{natbib}
\usepackage{helvet,soul}
\usepackage[font=small]{caption}

\begin{document}


\journalvol{60}
\journalyear{2018}
\journaleditors{P. Benaglia, A.C. Rovero, R. Gamen \& M. Lares}


\contriblanguage{0}


\contribtype{1}

\thematicarea{7}

\title{Oscilaciones de estrellas híbridas utilizando la aproximaci\'on de Cowling}


\titlerunning{Oscilaciones de estrellas híbridas utilizando la aproximación de Cowling}


\author{I.F. Ranea-Sandoval\inst{1,2}, M. Mariani\inst{1,2}, O.M. Guilera\inst{3,4}}
\authorrunning{Ranea-Sandoval et al.}


\contact{iranea@fcaglp.unlp.edu.ar}

\institute{ Grupo de Gravitaci\'on Astrof\'isica y Relatividad, Facultad de Ciencias Astron\'omicas y Geof\'isicas, \\Universidad Nacional de La Plata, Paseo del Bosque S/N (1900), La Plata, Argentina. \and CONICET, Godoy Cruz 2290, 1425 Buenos Aires, Argentina. \and Instituto de Astrof\'isica de La Plata, CONICET, Argentina. \and Grupo de Ciencias Planetarias, Facultad de Ciencias Astron\'omicas y Geof\'isicas, \\ Universidad Nacional de La Plata, Paseo del Bosque S/N (1900), La Plata, Argentina.}


\resumen{En el marco de la aproximaci\'on de Cowling, presentamos las frecuencias de los modos de oscilaci\'on $f$ (fundamental), $p_1$ (primer modo de presi\'on) y $g$ (gravitacional) de las perturbaciones cuadrupolares de objetos compactos construidos utilizando diferentes ecuaciones de estado. Pondremos especial atenci\'on a las llamadas estrellas h\'ibridas, compuestas por un n\'ucleo de materia de quarks y una envoltura de materia hadr\'onica. En este contexto, la existencia del modo $g$ ocurre \'unicamente de existir una transici\'on de primer orden entre la fase hadr\'onica y la de quarks.

}

\abstract{Within the framework of the Cowling approximation, we present the frequencies of the $f$ (fundamental), $p_1$ (first pressure) and $g$ (gravitational) of quadrupolar oscillation modes of compact objects constructed using different equations of state. Special attention will be put in hybrid stars, formed by a quark core and a hadronic envelope. In this context, $g$ modes are only present if a sharp first order phase transition occurs between hadronic and quark phases.

}



\maketitle

\section{Introducción}
\label{S_intro}
Recientemente se han detectado por primera vez ondas gravitacionales de modo directo provenientes de la fusión de agujeros negros de masas estelares \citep{GW2,GW1,GW3,GW4} y de estrellas de neutrones \citep{GW5} y su respectiva contraparte electromagnética \citep[ver, por ejemplo,][]{GW-EM}. Con estas detecciones no solo se confirma otra de las predicciones de la teoría general de la relatividad sino que se abre una nueva ventana observacional, la astronomía de ondas gravitacionales.

Las estrellas de neutrones son laboratorios astrofísicos. Estudios teóricos y observacionales mancomunados pueder permitirnos obtener información detallada del comportamiento de la materia sometida a condiciones extremas.

Si bien la detección de estrellas de neutrones con masas $\sim 2 M_\odot$ \citep{2solar1,2solar2} impone fuertes restricciones a la ecuación de estado de la materia dentro de una estrella de neutrones, todavía existen muchas capaces de reproducir estas observaciones \citep[ver, por ejemplo,][]{lattimerscience}. Mediciones precisas de las masas y los radios de este tipo de objetos permitirían comprender el comportamiento de la materia en el interior de estos objetos. Teniendo en cuenta que determinar el radio de este tipo de estrellas no es simple \citep[ver][y referencias allí mencionadas]{lattimer-steiner-2014}, la búsqueda de otros observables que permitan extraer información relacionada con la física de la materia en el interior de este tipo de estrellas resulta fundamental. Es en este punto que la detección de ondas gravitacionales provenientes de dichos objetos puede resultar de importancia central \citep[ver, por ejemplo,][]{GWfromNS,Lasky-2015}.

\section{Ecuaciones de estado y transiciones de fase}

\begin{figure*}[!t]
  \centering
  \includegraphics[width=0.4\textwidth]{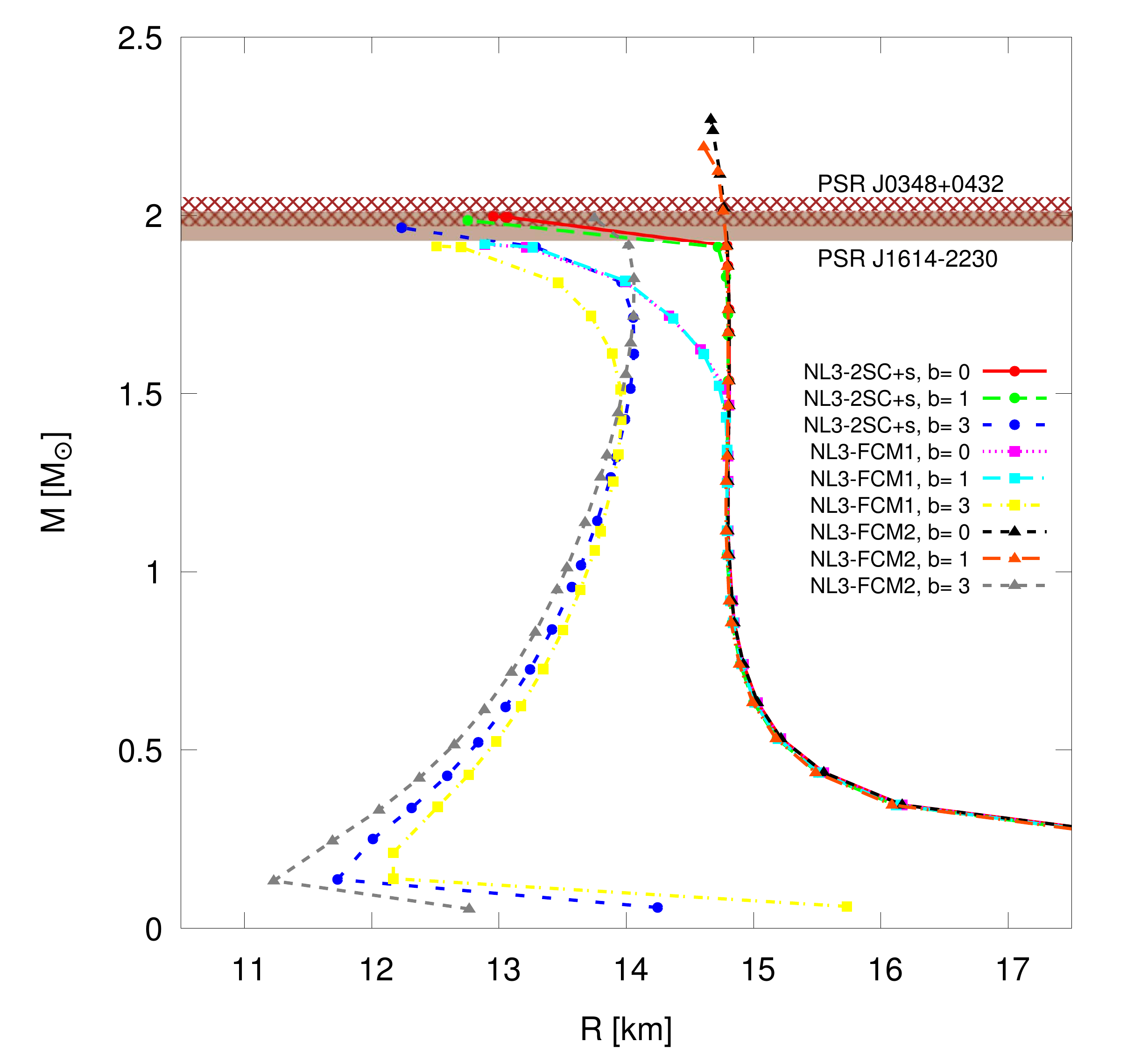} \includegraphics[width=0.4\textwidth]{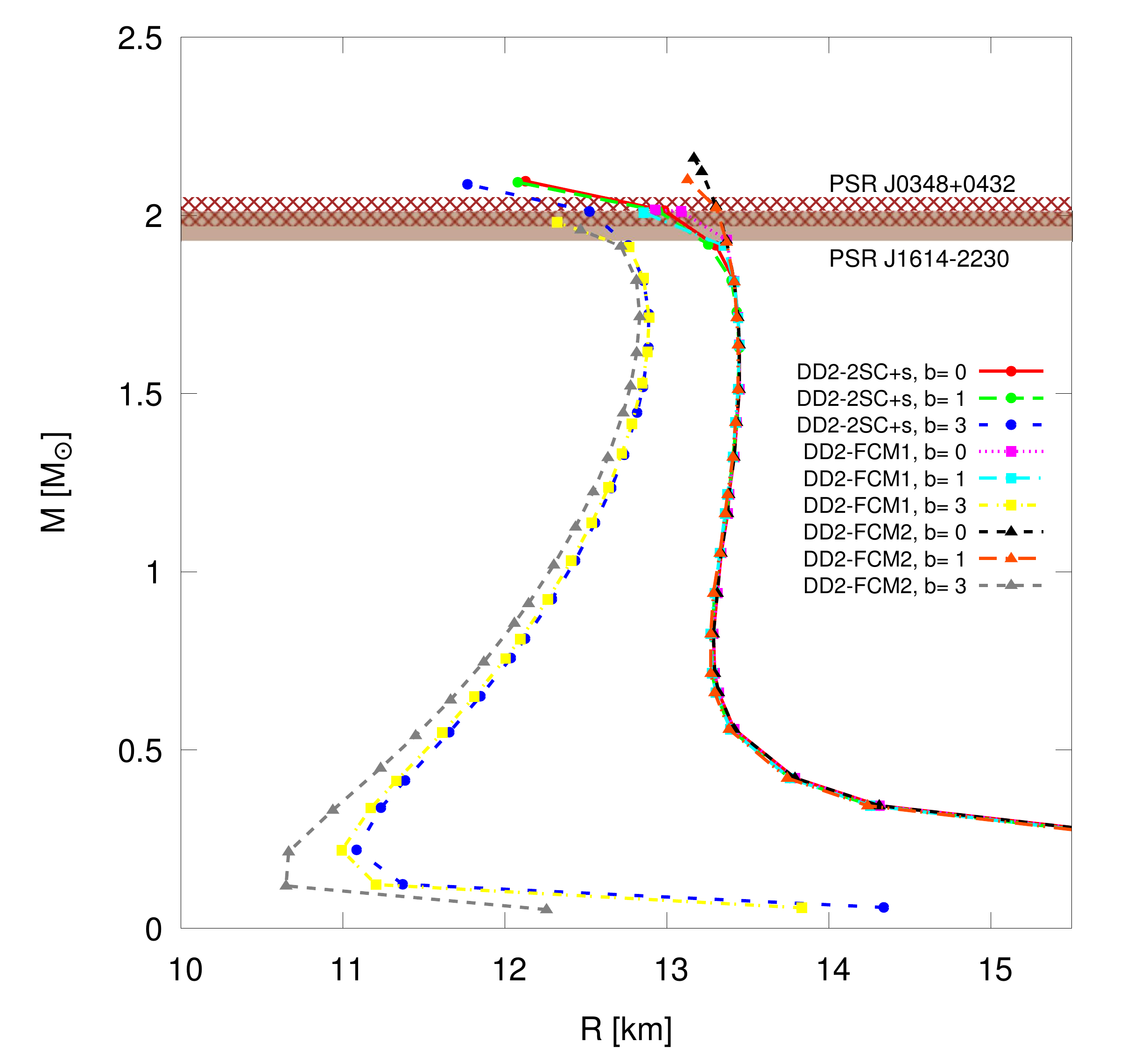}
  \caption{Panel izquierdo: Relación masa radio para diferentes EdE mixtas con parte hadrónica construida utilizado la parametrización NL3. Panel derecho: Idem pero utilizando la ecuación hadrónica DD2.
}
  \label{mraio}
\end{figure*}

Para este trabajo, construimos ecuaciones de estado (EdE) mixtas para modelar la materia en el interior de un objeto compacto. Para la corteza exterior utilizamos la famosa BPS desarrollada por \citet{Baym:1971,BAYM2}, para la corteza interior dos parametrizaciones diferentes de las conocidas aproximaciones relativistas de campo medio (RMF), la conocida como NL3 y la DD2, que incorpora constantes de acoplamiento dependientes de la densidad. Para el núcleo, donde asumimos la existencia de materia de quarks desconfinados, utilizamos una EdE basada en el Field Correlator Method (FCM) \citep[ver, por ejemplo,][]{FCM-b,FCM-a,Mariani.etal:2017} y otra, denominada 2SC+s, basada en un modelo NJL local \citep[ver, por ejemplo,][]{Orsaria:2013,Orsaria:2014}. La EdE del FCM queda caracterizada por dos parámetros: el potencial estático quark-antiquark asintótico, $V_1$, y el condensado gluónico, $G_2$. La EdE 2SC+s incorpora extrañeza y superconductividad de color en la fase 2SC y depende de los siguientes parámetros: la masa efectiva del quark $s$, las constantes de acoplamiento entre los diquarks, $\eta _{\rm qq}$, y la constante de acoplamiento vectorial, $\eta _{\rm v}$ \citep{RS-prep-2sc}.


La naturaleza de la transición de fases entre la materia de quarks y la hadrónica depende fuertemente del valor de la tensión superficial entre ambas fases, cantidad cuyo valor no está apropiadamente determinado \citep[ver][y referencias allí mencionadas]{Wu-shen-2017}. En este trabajo estudiaremos transiciones de fase abruptas en las que hay una discontinuidad en la densidad de energía (construcción de Maxwell) y presentaremos, también, una mirada heurística para la construcción de una fase mixta, en la que hadrones y quarks libres coexisten \citep[ver, por ejemplo,][]{Alvarez-Castillo2015}. Para construir esta fase mixta introduciremos funciones que interpolan suavemente por medio de aproximantes a la función escalón de Heaviside, donde el parámetro libre $b$ caracteriza el suavizado. El caso $b=0$ es equivalente a la construcción de Maxwell, mientras que valores mayores de $b$ indican mayores suavizados, lo cual resulta en regiones de fase mixtas más extendidas dentro de la estrella. Presentamos algunos ejemplos de ecuaciones de estado utilizadas en este trabajo en el panel izquierdo de la Figura~1, mientras que en el panel derecho de la misma presentamos las correspondientes relaciones masa-radio obtenidas al resolver las ecuaciones de equilibrio hidrostático relativistas, las ecuaciones de TOV.

\section{Oscilaciones en la aproximación de Cowling}

\begin{figure*}[!t]
  \centering
  \includegraphics[width=0.5\textwidth]{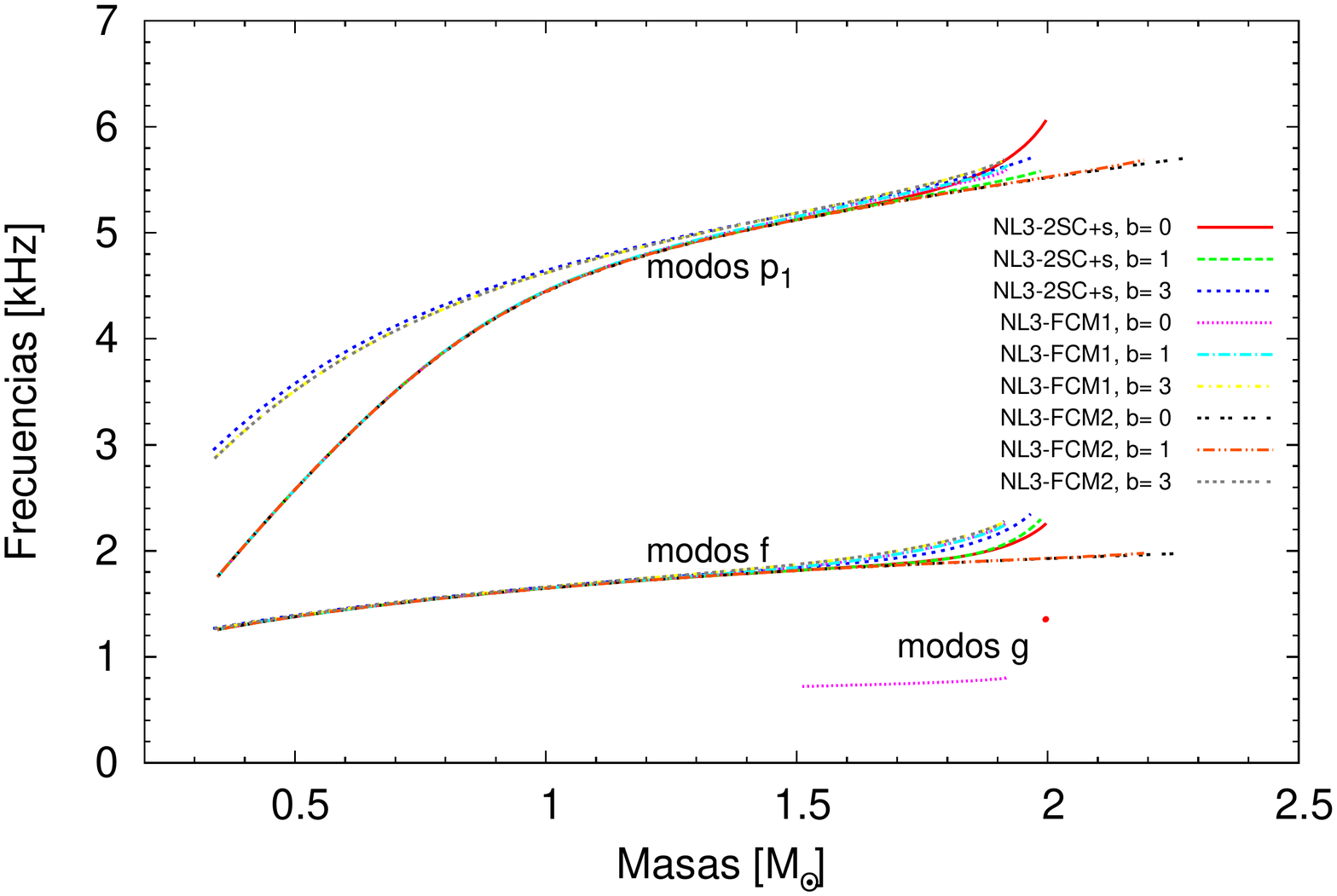}\includegraphics[width=0.5\textwidth]{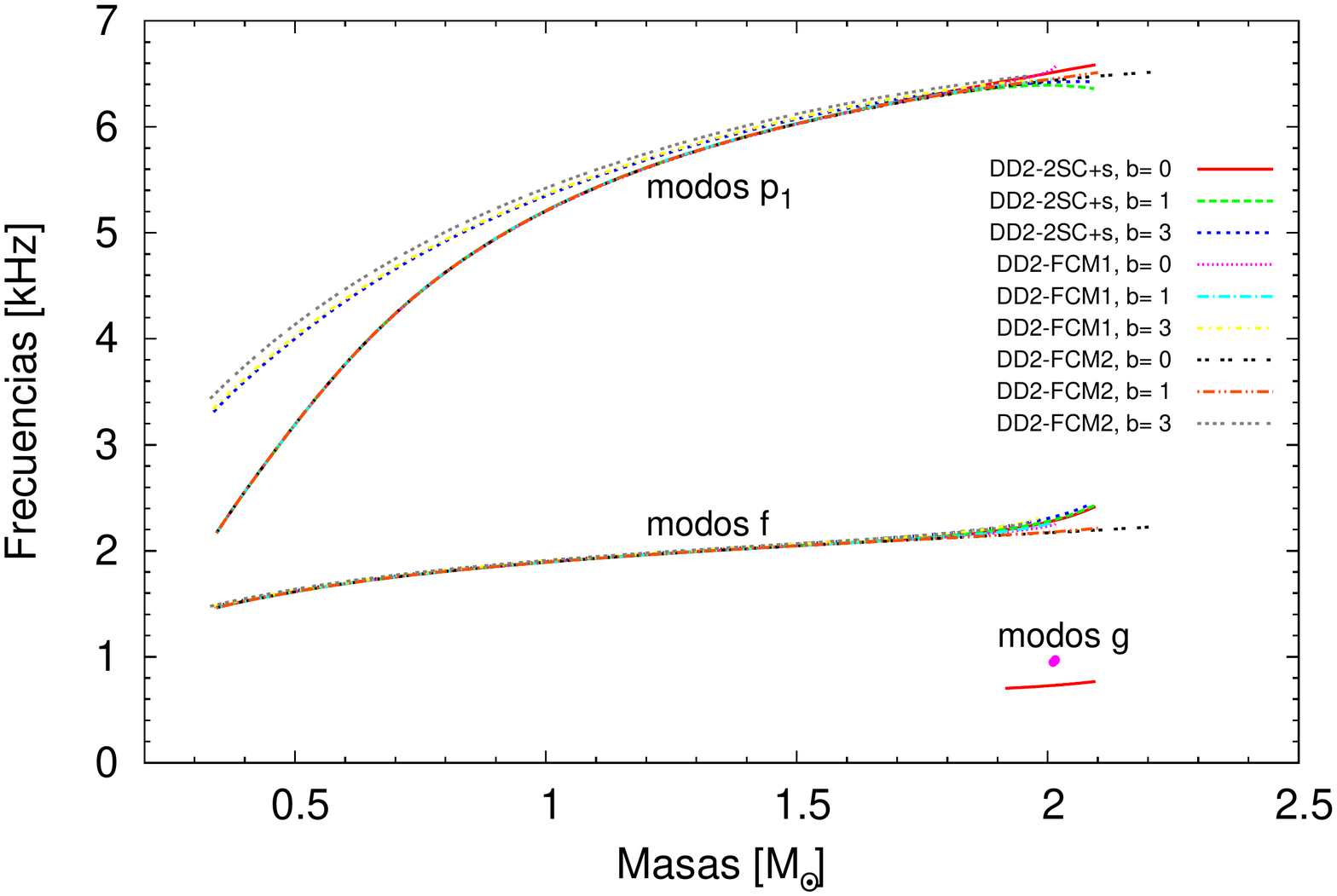}
  \caption{Panel izquierdo: Para la ecuación hadrónica NL3, se presentan las frecuencias de los modos $g$ (cuando están presentes), $f$ y $p_1$ para diferentes estrellas híbridas como función de la masa del objeto compacto. Panel derecho: Idem pero utilizando la ecuación hadrónica DD2.
}
  \label{modos}
\end{figure*}

El trabajo \citet{NS-oscillations} dio comienzo a los estudios relacionados con las oscilaciones de estrellas relativistas. Para estudiar los modos de oscilaciones del fluido ($g$, de gravedad, $f$, el fundamental y $p$, de presión), se puede utilizar la aproximación de Cowling \citep[ver, por ejemplo,][]{Finn:1988}. Con esta aproximación se obtienen valores para las frecuencias de oscilación que difieren en menos de un $20 \%$ con los obtenidos utilizando las ecuaciones linealizadas de la relatividad general \citep[ver][y referencias allí mencionadas]{vasquez-lugones:2013}. Es importante recordar que, en este formalismo, la aparición del modo $g$ es un indicador inequívoco de que la transición de fase entre materia hadrónica y de quarks es abrupta, ya que los mismos resultan completamente inhibidos si en la ecuación de estado no hay discontinuidades.

Para analizar las oscilaciones en el contexto de la aproximación de Cowling se debe, primero, a partir de la ecuación de estado, obtener la estructura de la estrella, resolviendo las ecuaciones de TOV. Sobre este fondo, se debe resolver un sistema de dos ecuaciones diferenciales ordinarias acopladas y satisfacer una condición de contorno en el radio de la estrella. Para las ecuaciones de estado que presentan discontinuidades se agrega, en el radio donde ocurre dicha transición, un par de condiciones adicionales que deben satisfacer las funciones que describen a las perturbaciones. Aquellas frecuencias que satisfacen este conjunto de condiciones de contorno son las de los modos propios de oscilación de la estrella. Para resolver la integración de las ecuaciones diferenciales acopladas, implementamos un código numérico que realiza la integración utilizando el método predictor-corrector de Runge-Kutta-Fehlberg. Para obtener las frecuencias que satisfacen las condiciones de borde, aplicamos un método correctivo tipo Newton-Raphson acoplado con el método de Ridders (dada las particularidades de las ecuaciones a resolver).

En el panel izquierdo de la Fig.~\ref{modos} presentamos, para perturbaciones cuadrupolares, las frecuencias de oscilación del modo fundamental, $f$, del primer modo de presión, $p_1$ y (cuando está presente) del modo de gravedad, $g$, como función de la masa del objeto compacto para diferentes ecuaciones de estado híbridas, utilizando para la parte hadrónica la parametrización NL3. Mientras que en el panel derecho de la Fig.~\ref{modos}, hacemos lo propio pero utilizando la ecuación de estado DD2.

Las diferentes familias de modos de oscilación están claramente diferenciadas entre sí. Los modos $g$ tienen frecuencias menores a $1$ kHz, los modos $f$ frecuencias $\sim 1$ kHz, mientras que el modo $p_1$ presenta frecuencias en el rango entre $2 - 6$ kHz.

Las curvas presentadas en la Fig.~\ref{modos}  muestran que no hay diferencias cualitativas entre las frecuencias obtenidas cuando se utiliza NL3 o DD2 como ecuación de estado para la parte hadrónica. También puede observarse que la aparición de materia de quarks en el núcleo de la estrella híbrida hace que las frecuencias del modo $f$ aumenten en relación a las de una estrella puramente construida con materia hadrónica.

\section{Conclusiones}

Presentamos cálculos de los modos de oscilación $g$, $f$ y $p_1$ de estrellas híbridas, utilizando la aproximación relativista de Cowling. Presentamos resultados obtenidos utilizando ecuaciones de estado híbridas construidas en base a las parametrizaciones NL3 y DD2 para la parte hadrónica y dos ecuaciones diferentes para la materia de quarks: el FCM y un modelo NLJ local con interacciones vectoriales que incorpora superconductividad de color en la fase 2SC. Para el tratamiento de la transición de fase, utilizamos la construcción de Maxwell y, además, presentamos un modelo heurístico para estudiar efectos asociados con la aparición de una fase mixta.

 Detecciones de ondas gravitacionales provenientes de estrellas de neutrones permitirían superar las dificultades asociadas con la determinación del radio de este tipo de estrellas. Además, la detección de modos $g$ serviría para arrojar luz sobre la naturaleza de la transición de fase entre la materia hadrónica y la de quarks: las estrellas de neutrones se vuelven verdaderos laboratorios astrofísicos.

\begin{acknowledgement}
Los autores quieren agradecer al Prof. Dr. Germán Lugones por sus aportes durante etapas iniciales de este trabajo y al Prof. Dr. Héctor Vucetich por mostrarnos las virtudes del método de Ridders. IFR-S y MM agradecen el aporte financiero de CONICET y la UNLP con los subsidios PIP-0714 y G140. OMG agradece el aporte financiero de CONICET y la UNLP con los subsidios PIP-0436 y G144.
\end{acknowledgement}


\bibliographystyle{baaa}
\small
\bibliography{IFRS}
 
\end{document}